\documentclass[preprint,showpacs,preprintnumbers,amsmath,amssymb,nofootinbib]{revtex4}
\usepackage{color}
\usepackage{booktabs}
\usepackage{mathrsfs}
\usepackage{epsfig}
\usepackage{graphicx}
\usepackage{dcolumn}
\usepackage{bm}
\usepackage{amsmath}
\usepackage{slashed}
\usepackage{multirow}

\let\jnfont=\rm
\def\NPB#1,{{\jnfont Nucl.\ Phys.\ B }{\bf #1},}
\def\PLB#1,{{\jnfont Phys.\ Lett.\ B }{\bf #1},}
\def\EPJC#1,{{\jnfont Eur.\ Phys.\ Jour.\ C }{\bf #1},}
\def\PRD#1,{{\jnfont Phys.\ Rev.\ D }{\bf #1},}
\def\PRL#1,{{\jnfont Phys.\ Rev.\ Lett.\ }{\bf #1},}
\def\MPLA#1,{{\jnfont Mod.\ Phys.\ Lett.\ A }{\bf #1},}
\def\JPG#1,{{\jnfont J.\ Phys.\ G}{\bf #1},}
\def\CTP#1,{{\jnfont Commun.\ Theor.\ Phys.\ }{\bf #1},}
\def\ZPC#1,{{\jnfont Z.\ Phys.\ C }{\bf #1},}
\def\JHEP#1,{{\jnfont JHEP \ }{\bf #1},}
\def\Rv{\not{\hbox{\kern-1pt $R$}}}
\def\p{\not{\hbox{\kern-3pt $p$}}}

\newcommand{\bea}{\begin{eqnarray}}
\newcommand{\eea}{\end{eqnarray}}

\newcommand{\bcen}{\begin{center}}
\newcommand{\ecen}{\end{center}}

\newcommand{\beq}{\begin{eqnarray}}
\newcommand{\eeq}{\end{eqnarray}}

\def\t1{\tilde{t_1}}

\begin{document}

\title{Implications of CP-violating Top-Higgs Couplings at LHC and Higgs Factories}
\author{ Archil Kobakhidze$^{1}$}
\author{ Ning Liu$^{1,2}$}
\author{ Lei Wu$^{1,3}$}
\author{ Jason Yue$^{1,4}$}
\affiliation{
$^1$ ARC Centre of Excellence for Particle Physics at the Terascale, School of Physics, The University of Sydney, NSW 2006, Australia\\
$^2$ Institution of Theoretical Physics, Henan Normal University, Xinxiang 453007, China\\
$^3$ Department of Physics and Institute of Theoretical Physics, Nanjing Normal University, Nanjing, Jiangsu 210023, China\\
$^4$Department of Physics, National Taiwan Normal University, Taipei 116, Taiwan
}%


\begin{abstract}
We utilise the LHC Run-1 and -2 Higgs data to constrain the CP-violating top-Higgs couplings. In order to satisfy the current full Higgs data sets at $2\sigma$ level, the CP-odd component $C^p_t$ and the CP-even component $C^s_t$ have to be within the ranges $|C^p_t|< 0.37$ and $0.85 <C^s_t< 1.20$, respectively, which is stronger than the LHC Run-1 bound $|C^p_t|< 0.54$ and $0.68 <C^s_t< 1.20$. With the new bound, we explore the impact on the CP-violating top-Higgs couplings in the Higgs production processes $pp \to t\bar{t}h, thj, hh$ at 13 TeV LHC, and $e^+e^- \to hZ, h\gamma$ at future 240 GeV Higgs factories. We find that the cross sections of $pp \to t\bar{t}h$, $pp \to thj$, $pp \to hh$, $e^+e^- \to hZ$ and $e^+e^- \to h\gamma$ can be enhanced up to 1.41, 1.18, 2.20, 1.001 and 1.09 times as large as the SM predictions, respectively. The future precision measurement of the process $e^+e^- \to h\gamma$ with an accuracy of 5\% will be able to constrain $|C^p_t|<0.19$ at most at a 240 GeV $e^+e^-$ Higgs factory.

\end{abstract}

\pacs{12.60.Jv, 14.80.Ly}
\maketitle

\section{INTRODUCTION}
In the Run-1 of LHC, the existence of the Higgs boson was firmly established \cite{higgs-atlas,higgs-cms}. The signal was observed with large significance in four decay channels ($\gamma\gamma$, $ZZ^*$, $WW^*$, $\tau\tau$) and two main production modes (gluon fusion, vector-boson fusion) have been isolated. Currently the LHC is running at a center-of-mass energy of 13 TeV. With the available $\sim 13$ fb$^{-1}$ data, the ATLAS and CMS experiments updated their measurements of the Higgs boson at LHC Run-2. The measured Higgs properties are compatible with the predictions in the Standard Model (SM). Still, due to the large uncertainties, measurements of several Higgs couplings such as that to the top quark, is required to be refined.

Within the Standard Model, the value of the top-Higgs coupling is known to a good accuracy, as it is proportional to the mass of the top quark. Yet, such proportionality can be distorted in models beyond the SM, for example, the non-linear realisation of the electroweak gauge symmetry \cite{Callan:1969sn,Coleman:1969sm}. Moreover, top-Higgs coupling plays an important role in vacuum stability \cite{Sher:1988mj,Degrassi:2012ry} and  electroweak baryogenesis \cite{Zhang:1994fb,Kobakhidze:2015xlz,Huang:2015izx}. Thus, measurements of the top-Higgs coupling provide a crucial  clues to new physics.

The top-Higgs coupling contributes to the main Higgs production channel $gg \to h$ and the clean Higgs decay channel $h \to \gamma\gamma$ through the quantum effects at the LHC. These rates are measured to well agree with the SM, therefore serves as strong constraints on the top-Higgs coupling \cite{Kobakhidze:2012wb,Cheung:2014noa,Kobakhidze:2014gqa,Dolan:2014upa,Khatibi:2014bsa,Chien:2015xha,Cirigliano:2016njn,Cirigliano:2016nyn}. On the other hand, the top-Higgs coupling can affect other important Higgs processes, such as the Higgs pair production at the LHC \cite{Dolan:2012ac,Nishiwaki:2013cma,Han:2013sga,Liu:2014rba,Dawson:2015oha,Goertz:2014qta,Shen:2015pha,Wu:2015nba,Lu:2015jza,Cao:2015oaa}. Therefore, it is meaningful to see how much room is left for the anomalous top-Higgs coupling and its impact on various Higgs processes within the allowed parameter space.

In this work, we perform an updated fit of the most general top-Higgs coupling to the Higgs data collected by LHC Run-1 and -2. Using the bounds on top-Higgs coupling, we examine the effect of anomalous top-Higgs coupling on the production rates of the Higgs processes $pp \to t\bar{t}h,thj,hh$ at the LHC and $e^+e^- \to hZ,h\gamma$ at future lepton colliders, such as ILC, FCC-ee and CEPC.
\begin{itemize}
\begin{figure}[ht]
\centering
\includegraphics[width=4in,height=1.5in]{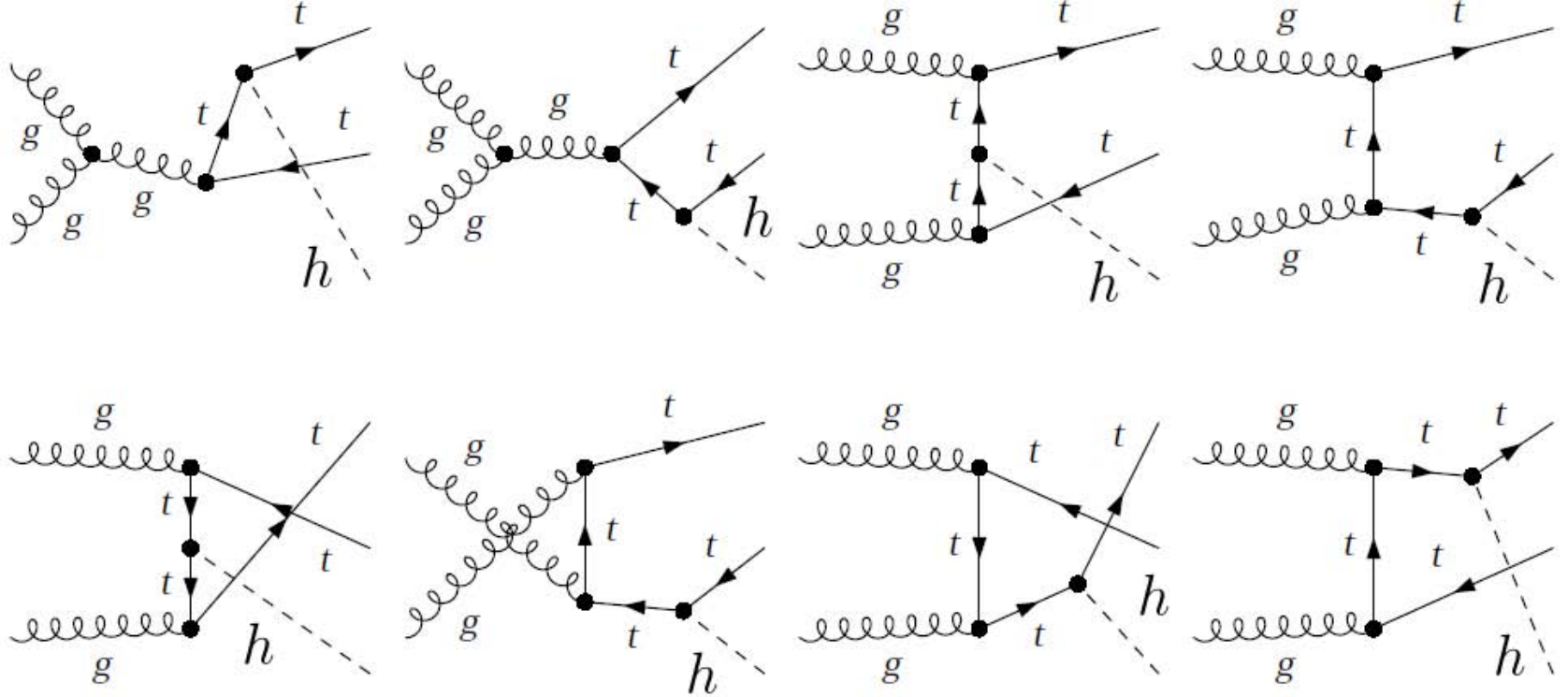}\vspace{-0.5cm}
\caption{Representative parton-level Feynman diagrams of the process $pp \to t\bar{t}h$  in leading order at the LHC.}
\label{fig:tth}
\end{figure}
\begin{figure}[ht]
\centering
\includegraphics[width=4in,height=1in]{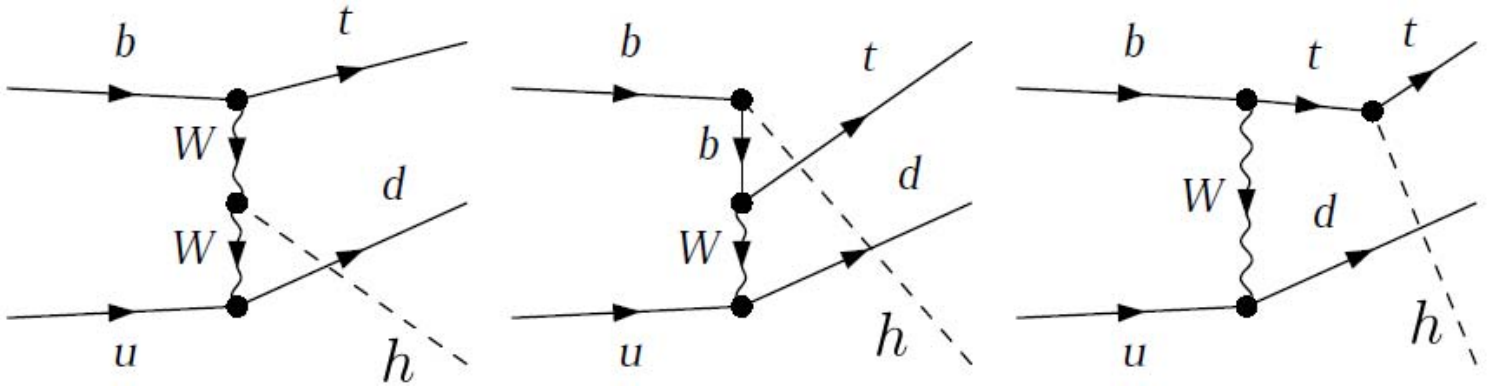}\vspace{-0.5cm}
\caption{Representative parton-level Feynman diagrams of the process $pp \to thj$ in leading order at the LHC.}
\label{fig:thj}
\end{figure}
  \item The direct measurements of top-Higgs coupling come mainly from the associated production of the top pair with Higgs boson and the single top associated production with the Higgs boson at the LHC. The former has the larger production rate \cite{Marciano:1991qq,Frederix:2011zi,Degrande:2012gr,Liu:2015aka,Buckley:2015vsa,Cao:2016wib,Maltoni:2016yxb,Gritsan:2016hjl,Dolan:2016qvg,Chang:2016mso}, but the latter can be used to determine the sign of top-Higgs coupling \cite{Kobakhidze:2014gqa,Maltoni:2001hu,Lu:2010zzb,Farina:2012xp,Biswas:2012bd,Ellis:2013yxa,Englert:2014pja,Chang:2014rfa,Wu:2014dba,Yang:2014xma,Yue:2014tya,Rindani:2016scj,Liu:2016dag}.
  \item The di-Higgs production is the only way to measure the Higgs self-coupling at the LHC, which is dominated by the gluon-gluon fusion mechanism. The cross section of di-Higgs production is $\sim\mathcal{O}(10^3)$ smaller than the single Higgs production at the LHC. This is because that there is a strong cancelation between the box diagrams from top-Higgs coupling and the triangle diagrams from the Higgs self-coupling \cite{Kniehl:1995tn}. So, any changes in top-Higgs coupling will affect the determination of the Higgs self-coupling at the LHC.
\begin{figure}[ht]
\centering
\includegraphics[width=5in,height=2in]{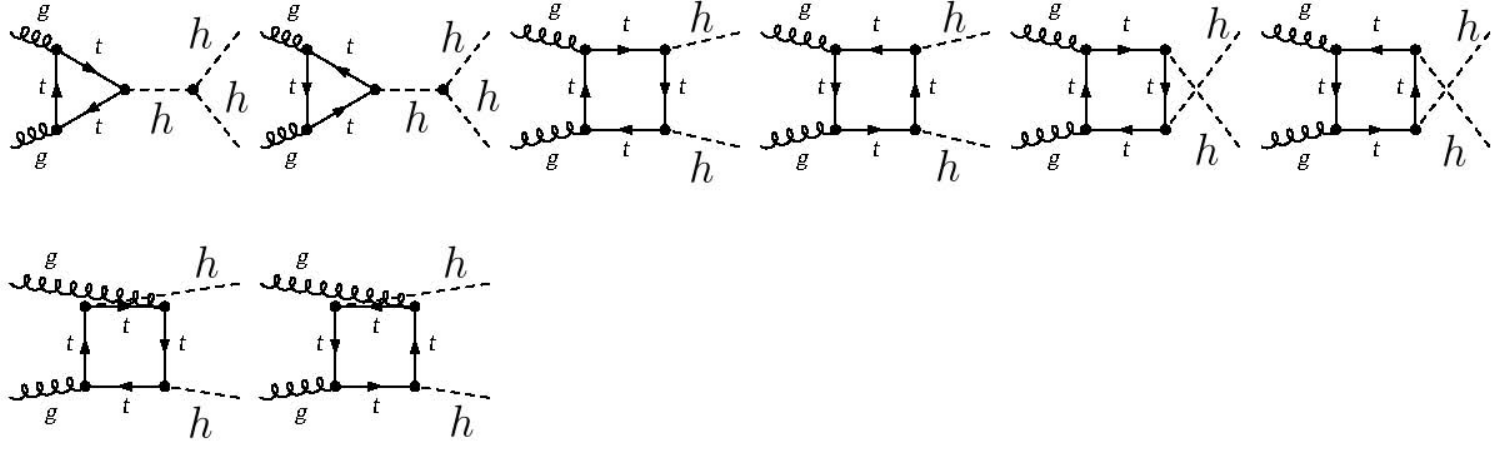}\vspace{-0.5cm}
\caption{Representative parton-level Feynman diagrams of the process $pp \to hh$ in leading order at the LHC.}
\label{fig:hh}
\end{figure}
  \item The top-Higgs coupling also contributes to the main Higgs production process, $e^+e^- \to hZ$, via loops at future Higgs factories. The precision measurement of the production rate of $e^+e^- \to hZ$ therefore possibly constrain the top-Higgs coupling. Another interesting process is $e^+e^- \to h\gamma$ \cite{Hu:2014eia,Ren:2015uka,Cao:2015iua}, which appears at one-loop level in the SM. Such a process is found to be a sensitive probe in testing the CP nature of the top-Higgs coupling since the CP-violating top-Higgs couplings can induce a large forward-backward asymmetry \cite{Li:2015kxc}.
\begin{figure}[ht]
\centering
\includegraphics[width=5.5in,height=1.0in]{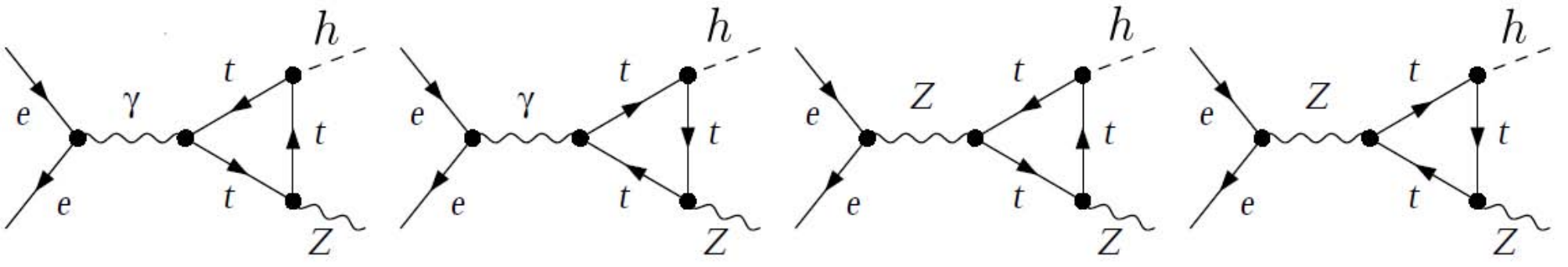}\vspace{-0.5cm}
\caption{Representative Feynman diagrams of the one-loop virtual corrections to the process $e^+e^- \to hZ$  in next-to-leading order at Higgs Factories.}
\label{fig:hz}
\end{figure}
\begin{figure}[ht]
\centering
\includegraphics[width=5.5in,height=1.0in]{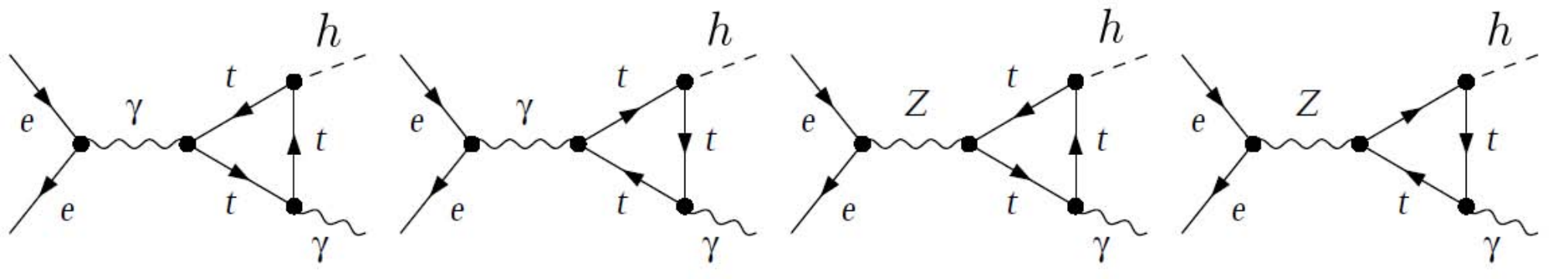}\vspace{-0.5cm}
\caption{Representative Feynman diagrams of the process $e^+e^- \to h\gamma$ in leading order at Higgs Factories.}
\label{fig:hr}
\end{figure}
\end{itemize}

The structure of this paper is organised as follows. In Section \ref{section2}, we will briefly introduce the non-standard top-Higgs interaction and the relevant constraints. In Section \ref{section3}, we present the numerical results and discuss the effects of non-standard top-Higgs coupling in the Higgs production processes $pp \to t\bar{t}h, thj, hh$ at 13 TeV LHC, and $e^+e^- \to hZ, h\gamma$ at future 240 GeV Higgs factories. Finally, the conclusions are drawn in Section \ref{section4}.

\section{Constraints on Top-Higgs interaction}\label{section2}
In the SM, the top-Higgs coupling is a purely scalar interaction. However, in models beyond the SM, such as the non-linear realisation of the electroweak gauge symmetry, the top-Higgs coupling can comprise both scalar and pseudoscalar components. The most general form of the top-Higgs coupling can be parameterised as follows:
\begin{equation}
{\cal L} \supset -\frac{m_t}{v}\overline{t}(C^s_t+iC^p_t\gamma^{5})th,
\end{equation}
where $m_t$ is the top quark mass and $v$ is the vacuum expectation value of the Higgs field. In the SM, $C^s_t=1$ and $C^p_t=0$ at leading order. But the $\mathcal{CP}$-violating component $C^p_t$ can arise from the high order corrections, whose value is expected to be small \cite{AguilarSaavedra:2009mx}.

\begin{figure}[ht]
\centering
\includegraphics[width=.35\textwidth]{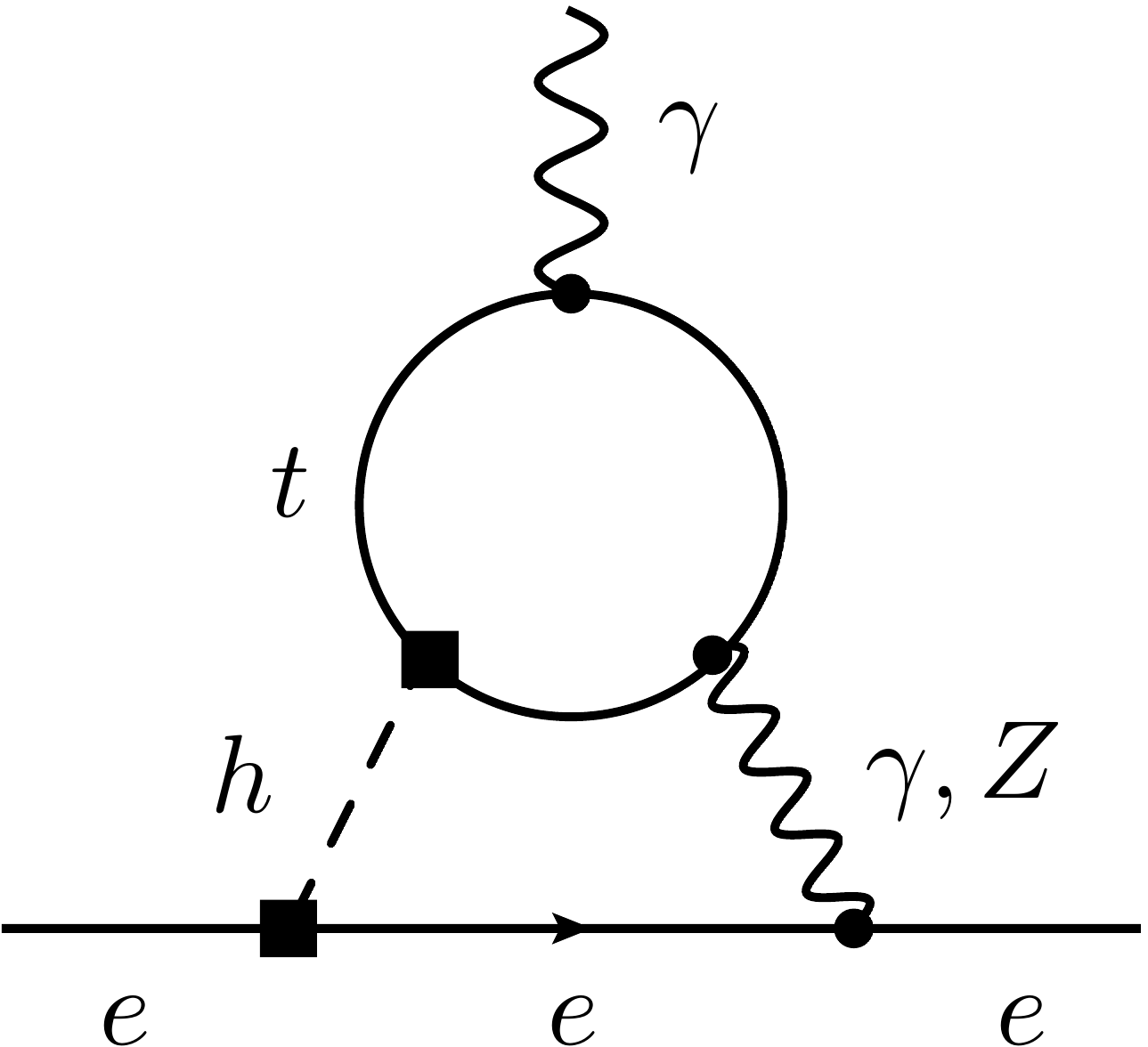}\vspace{-0.5cm}
\caption{Two-loop Barr-Zee contributions to the electron EDM.}
\label{br}
\end{figure}
In general, the $\mathcal{CP}$-violating interactions can contribute to the electron, neutron and mercury electric dipole moments (EDMs). Among them, the electron EDM (eEDM) produces the strongest bound on $C^p_t$ \cite{Brod:2013cka}. The dominant contribution to the eEDM comes from the diagram with the exchange of photon, which is given by \cite{Barr:1990vd}:
\begin{equation} \label{first}
\frac{d_e}{e} = \frac{16}{3} \frac{\alpha}{(4\pi)^3} \hspace{0.25mm}
\sqrt{2} G_F  \hspace{0.25mm}  m_e \, \Big[  C^s_e C^p_t
  \, f_1 (x_{t/h}) +  C^p_e C^s_t  \, f_2 (x_{t/h}) \Big
] \,,
\end{equation}
where $x_{t/h} := m_t^2/m_h^2$ and the loop functions $f_{1,2} (x)$ can be written as~\cite{Stockinger:2006zn}:
\begin{equation}
\begin{split}
f_1 (x) & = \frac{2x}{\sqrt{1-4x}} \left [ {\rm Li}_2  \left ( 1 - \frac{1- \sqrt{1-4x}}{2x} \right ) - {\rm Li}_2  \left (  1 -
  \frac{1+ \sqrt{1-4x}}{2x} \right )  \right ] \,, \\[4mm]
f_2 (x) & = \left (1 - 2 x \right) f_1(x) + 2 x \left (\ln x + 2 \right)   \,.
\end{split}
\end{equation}
Here ${\rm Li}_2 (x) = -\int_0^x du \, \ln (1-u)/u$ is the usual dilogarithm. The ACME limit reads \cite{Baron:2013eja}:
\begin{equation} \label{eq:bestde}
\left | \frac{d_e}{e} \right | < 8.7 \cdot 10^{-29} \, {\rm cm} \,,
\end{equation}
However, as shown in Eq.(\ref{first}), such a bound on the coupling $C^p_t$ depend on the assumption of Higgs couplings to the electron, which can vanish by an appropriate tuning of the ratio $C^p_e/C^s_e$. Since the electron-Higgs couplings are practically unobservable at the LHC, we do not impose EDM constraint in this work. The indirect constraints from flavor physics, such as $B_s-\overline{B}_s$ mixing and $B \to X_s \gamma$, remain relatively weak, due to the theoretical uncertainties \cite{Brod:2013cka}.

\begin{table}[t!]
\caption{
The best-fit values for various Higgs signal strengths $\mu$ from the CMS and ATLAS 13 TeV experiments, respectively. }
\medskip
\begin{ruledtabular}
\begin{tabular}{lcc}
%
Category & ${\rm ATLAS}$ & ${\rm CMS}$   \\
\hline
 ggF$(\gamma\gamma) $ &  $ +0.62\,^{+0.30}_{-0.29} $~\cite{ATLAS-CONF-2016-081}  &  $ +0.77\,^{+0.25}_{-0.23} $~\cite{CMS-PAS-HIG-16-020}\\
 ggF$(4\ell) $        &  $ +1.34\,^{+0.39}_{-0.33} $~\cite{ATLAS-CONF-2016-081}  &  $ +0.96\,^{+0.40}_{-0.33} $~\cite{CMS-PAS-HIG-16-033} \\
 VBF$(\gamma\gamma) $ &  $ +2.25\,^{+0.75}_{-0.75} $~\cite{ATLAS-CONF-2016-081}  &  $ +1.61\,^{+0.90}_{-0.80} $~\cite{CMS-PAS-HIG-16-020} \\
 $V_{had}H(\gamma\gamma)$ & $ -0.81\,^{+2.20}_{-1.88} $~\cite{ATLAS-CONF-2016-081} & -- \\
 $V_{lep}H(\gamma\gamma)$ & $ +0.68\,^{+1.71}_{-1.30} $~\cite{ATLAS-CONF-2016-081} & --  \\
$t\bar{t}H_{combine}$ &  $ +1.8\,^{+0.7}_{-0.7} $ ~\cite{ATLAS-CONF-2016-068}  &  --  \\
$ t\bar{t}H(\gamma\gamma) $ &  -- &  $ +1.91\,^{+1.5}_{-1.2} $~\cite{CMS-PAS-HIG-16-020}\\
$ t\bar{t}H(3\ell) $ &  -- &  $ +1.3\,^{+1.2}_{-1.0} $~\cite{CMS-PAS-HIG-16-022} \\
$ t\bar{t}H(ss2\ell) $ &  -- &  $ +2.7\,^{+1.1}_{-1.0} $~\cite{CMS-PAS-HIG-16-022}
\label{data}
\end{tabular}
\end{ruledtabular}
\end{table}
Besides, the CP-violating top-Higgs interaction can affect the production rate of $gg \to h$ and decay width of $h \to \gamma\gamma$. At the LO, the Higgs rates normalised to the SM expectations can be written as:
\beq
\frac{\Gamma( h \to \gamma\gamma)}{\Gamma( h \to \gamma\gamma)\vert_{\rm SM}}
&  \simeq &
\frac{\big \vert \frac{1}{4} A_1[m_W] + (\frac{2}{3})^2 C^s_t \big \vert^2
+ \vert (\frac{2}{3})^2 \frac{3}{2} C^p_t) \vert^2}
{\big \vert \frac{1}{4} A_1[m_W] + (\frac{2}{3})^2  \big \vert^2 }
\nonumber \\
\frac{\sigma( gg \to h)}{\sigma( gg \to h)\vert_{\rm SM}} & = &
\frac{\Gamma( h \to gg)}{\Gamma( h \to gg)\vert_{\rm SM}}  \simeq
\big \vert C^s_t  \big \vert^2 + \vert \frac{3}{2} C^p_t
\big \vert^2
\label{Eq:widthsCP}
\eeq
with $A_1[m_W] \simeq -8.3$ for $m_h \approx 125$ GeV. From Eq.~\ref{Eq:widthsCP}, we can see that the pseudo-component $C^p_t$ of $t\overline{t}h$ has a larger contribution to $\sigma(gg \to h)$ than the scalar-component $C^s_t$. When $C^s_t$ increases, the ratio of $\sigma(gg \to h)/\sigma(gg \to h)\vert_{\rm SM}$ becomes larger. However, this observation is not applicable to the ratio $\Gamma( h \to \gamma\gamma)/\Gamma( h \to \gamma\gamma)\vert_{\rm SM}$, because the contribution of $C^s_t$ will cancel with $W$ boson loop.

The database \textsf{15.09} of \textsf{Lilith-1.1.3} \cite{Bernon:2015hsa} have been updated with the new Higgs data listed in Tab.~\ref{data} in the fit, which was performed under the reduced coupling mode. The total number of degrees of freedom, $n_{dof}$, is given by the differnce between the number of observables, $n_{obs}$, and the number of scan parameters, $n_{para}$. In our fit, we have in total $n_{obs} = 48~(35)$ and $n_{para} = 2$ after (before) LHC Run-2, corresponding to $n_{dof}=46~(33)$.

\begin{figure}[ht]
\centering
\includegraphics[width=3.2in,height=3in]{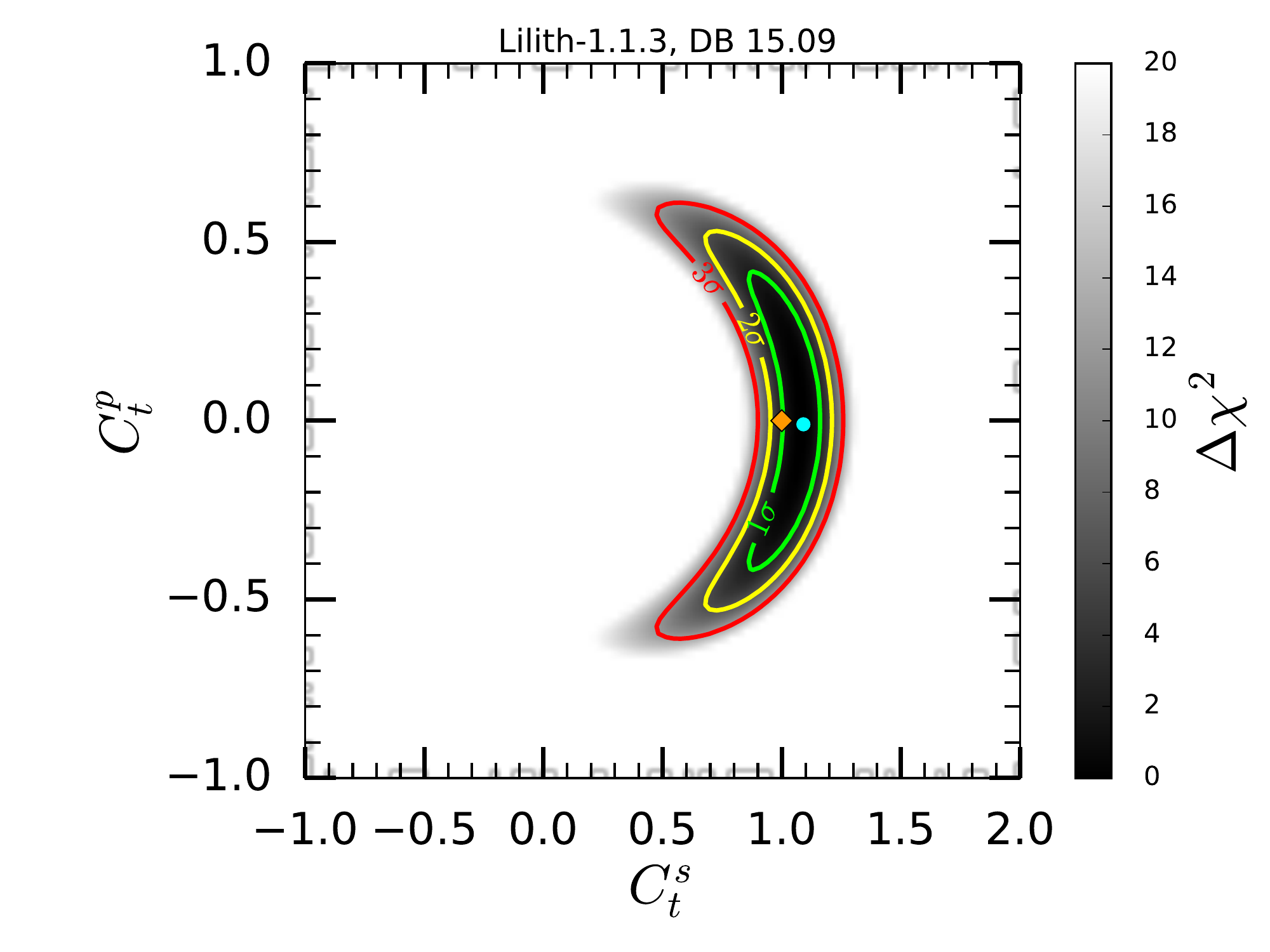}
\includegraphics[width=3.2in,height=3in]{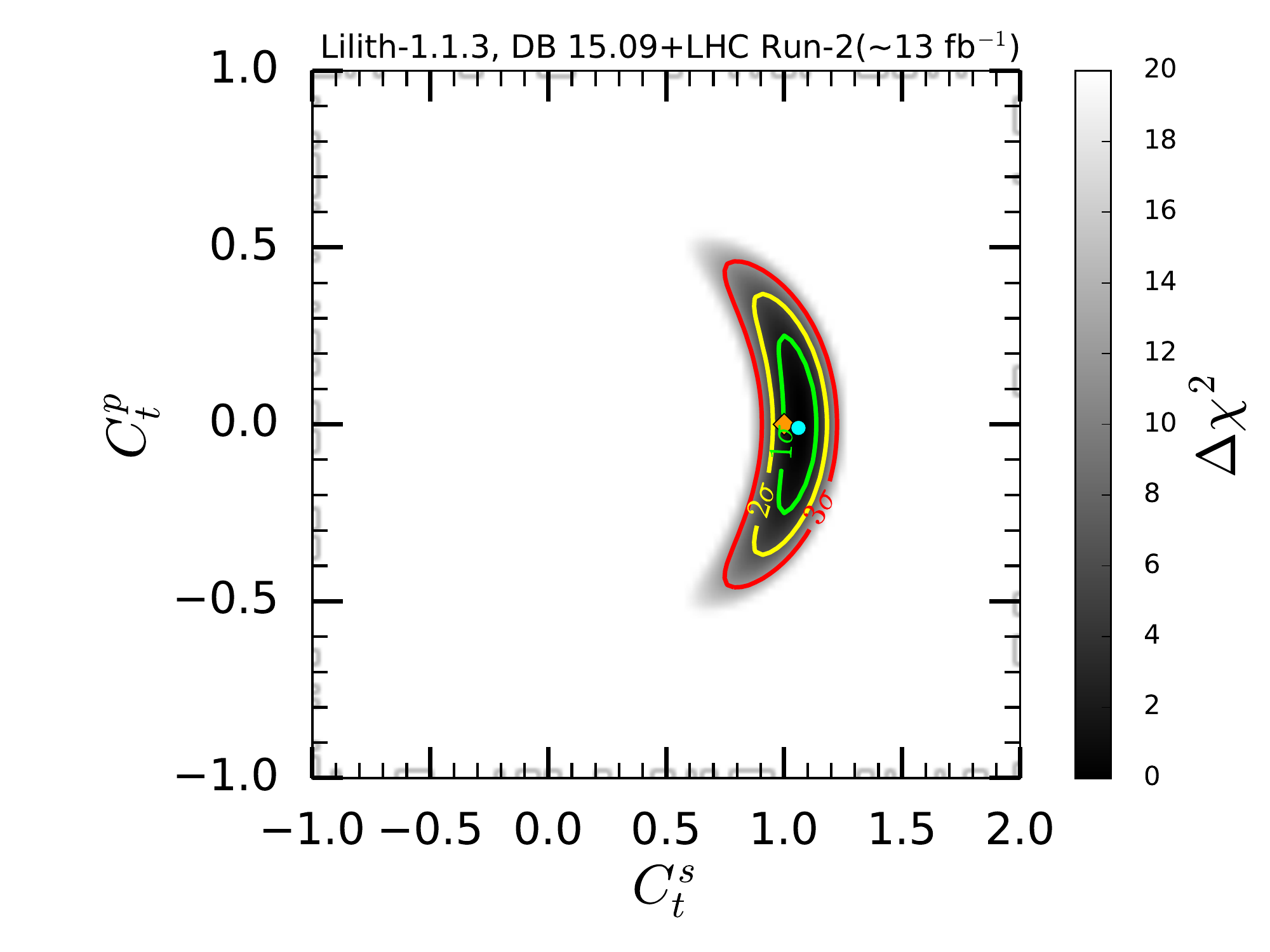}\vspace{-0.5cm}
\caption{Higgs data fit results of $C^s_t$ and $C^p_t$ for LHC Run-1 (left panel) and Run-1 with Run-2 (right panel). The orange diamond and cyan bullet correspond to the SM value of top-Higgs coupling and the best fit value, respectively. The color map is the value of $\Delta\chi^2$. }
\label{fit}
\end{figure}
In Fig.~\ref{fit}, we show the Higgs data constraints on the CP-violating top-Higgs couplings after (before) LHC Run-2. All other Higgs couplings are assumed to take the SM values. Points with $\Delta \chi^2 < 2.30$ (6.18, 11.83) correspond to points in a two-dimensional 1$\sigma$ ($2\sigma$, $3\sigma$) region in the Gau\ss ian limit. The left panel shows the bounds due to the LHC Run-1 data sets ($\sim 25$ fb$^{-1}$, $7+8$ TeV). The best fit values of $C^p_t$ and $C^s_t$ give $\chi^2_{min}/n_{dof}=28.96/33$. The bounds $|C^p_t|< 0.54$ and $0.85 <C^s_t< 1.2$ are required to be consistent with the Higgs data at $2\sigma$ level. In the right panel, the LHC Run-1 data sets have been combined with the available LHC Run-2 data sets ($\sim 13$ fb$^{-1}$, 13 TeV), which is shown in Table.~\ref{data}. The stronger $2\sigma$ limits of $|C^p_t|< 0.37$ and $0.68 <C^s_t< 1.2$ are obtained by using the combined data. This is mainly because that the current signal strengthes of diphoton final states in the gluon fusion channel is less than unity, which favors smaller values of $C^p_t$. The best fit values of $C^p_t$ and $C^s_t$ give $\chi^2_{min}/n_{dof}=41.90/46$. Negative values of $C^s_t$ are excluded in both fits.

\section{collider implications}\label{section3}
Given the importance of the top-Higgs coupling in measuring the properties of the Higgs boson, we examine its impact on the production rates of the processes $pp \to t\bar{t}h,~thj,~hh$ at the LHC and $e^+e^- \to hZ,~h\gamma$ at future lepton colliders within the allowed parameter ranges of $C^s_t$ and $C^p_t$.

In the numerical calculations, we take the input parameters of the SM as \cite{pdg}:
\begin{eqnarray}
m_t=173.07{\rm ~GeV}, ~~m_W = 80.385~, ~~m_{Z}=91.19 {\rm
~GeV},\nonumber
\\m_e = 0.519991 {\rm ~MeV},~~\sin^{2}\theta_W=0.2228, ~~\alpha(m_Z)^{-1}=127.918.
\end{eqnarray}
The value of the Higgs mass is taken as $m_h=125$ GeV. For the strong coupling constant $\alpha_s(\mu)$, we use the 2-loop evolution with the QCD parameter $\Lambda^{n_{f}=5}=226{\rm ~MeV}$. The CTEQ6M parton distribution functions (PDF) \cite{cteq} are chosen for the calculations. We set the renormalisation scale $\mu_R$ and factorisation scale $\mu_F$ to be $\mu_R=\mu_F=(\Sigma m_f)/2$. All the amplitudes are generated by \textsf{FeynArts-3.9} \cite{feynarts}, and are further reduced by \textsf{FormCalc-8.3} \cite{formcalc}. The loop functions are numerically calculated by \textsf{LoopTools-2.8} \cite{looptools}. We keep the electron mass and checked that all the UV divergences in the one-loop processes $e^+e^ \to hZ$ and $e^+e^ \to h\gamma$ are canceled. The infrared (IR) divergences can appear in the virtual correction of $e^+e^- \to hZ$, which is removed with the two cutoff phase space slicing method \cite{two-cutoff} used in our previous works \cite{ningliu-1,ningliu-2,ningliu-3}. To show the impact of the CP-violating top-Higgs couplings, we define the signal strength $\mu$ as,
\begin{eqnarray}
\mu_i=\frac{\sigma^{CPV}_i}{\sigma^{SM}_i}
\end{eqnarray}
where $i\in\{t\bar{t}h,~thj,~hh,~hZ,~h\gamma\}$. The advantage of this ratio is that it has a weak dependence on the renormalization scale.

\subsection{$t\bar{t}h$, $thj$ and $hh$ production at LHC}
\begin{figure}[h]
\centering
\includegraphics[width=2.in,height=2.5in]{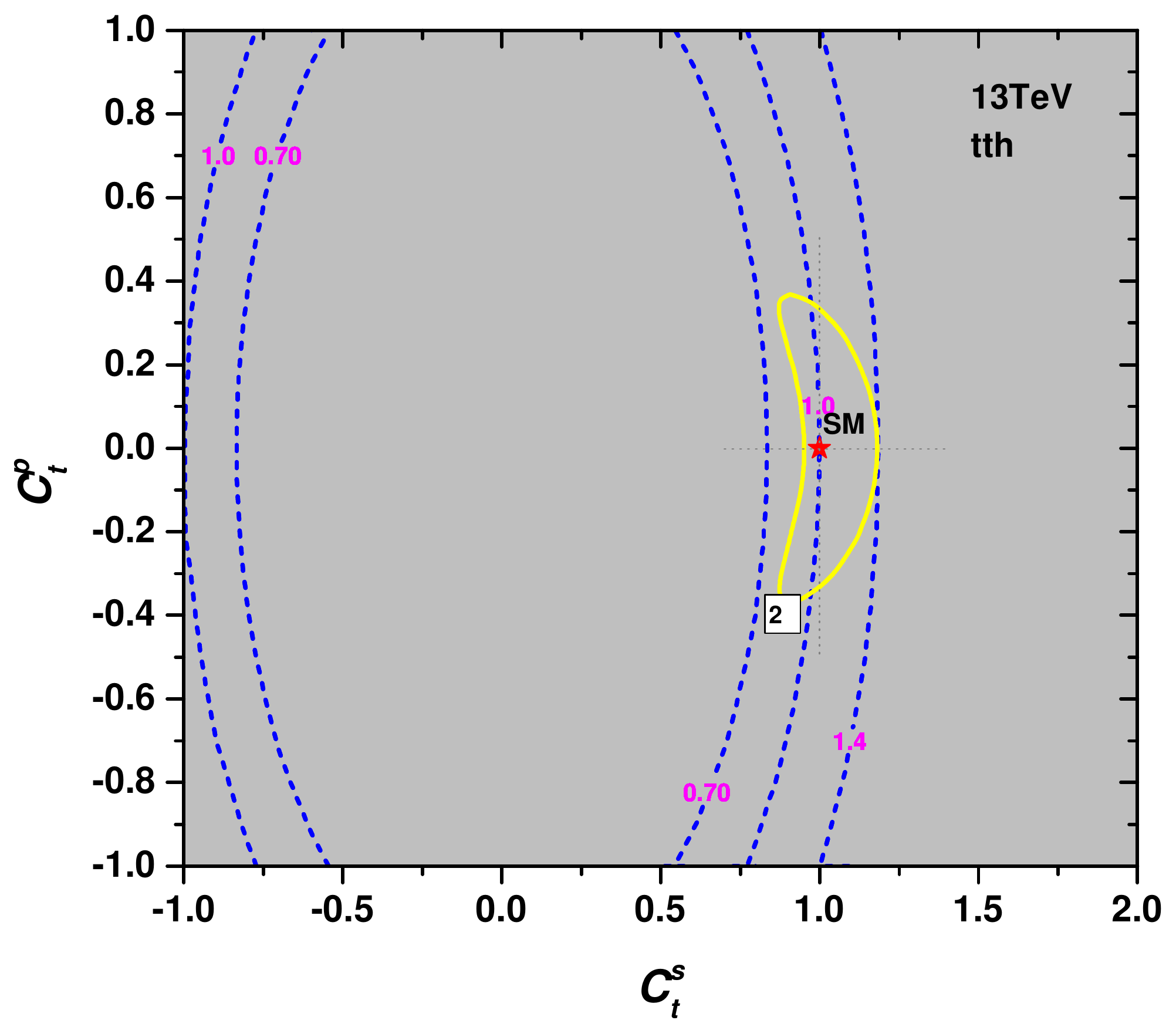}
\includegraphics[width=2.in,height=2.5in]{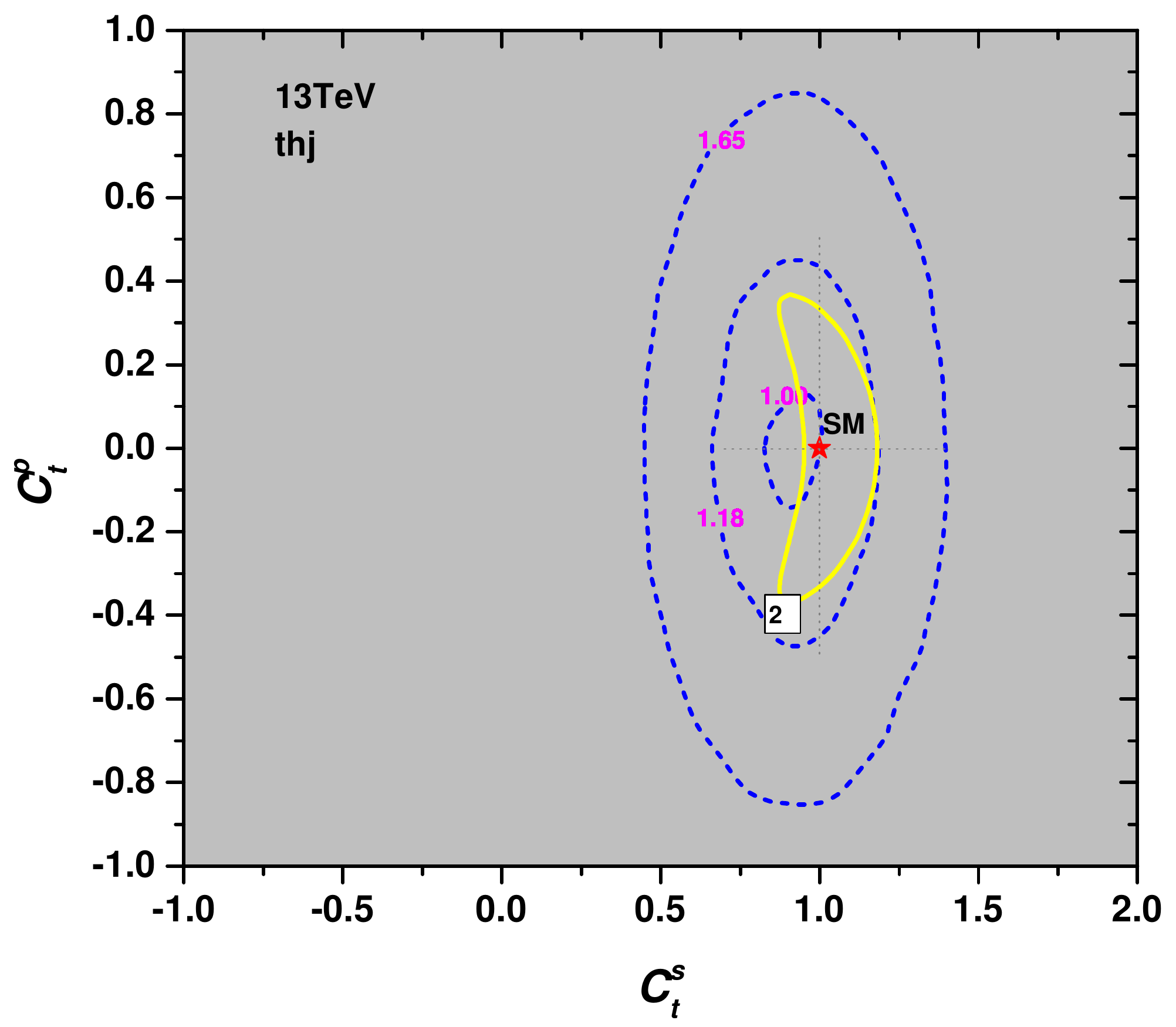}
\includegraphics[width=2.in,height=2.5in]{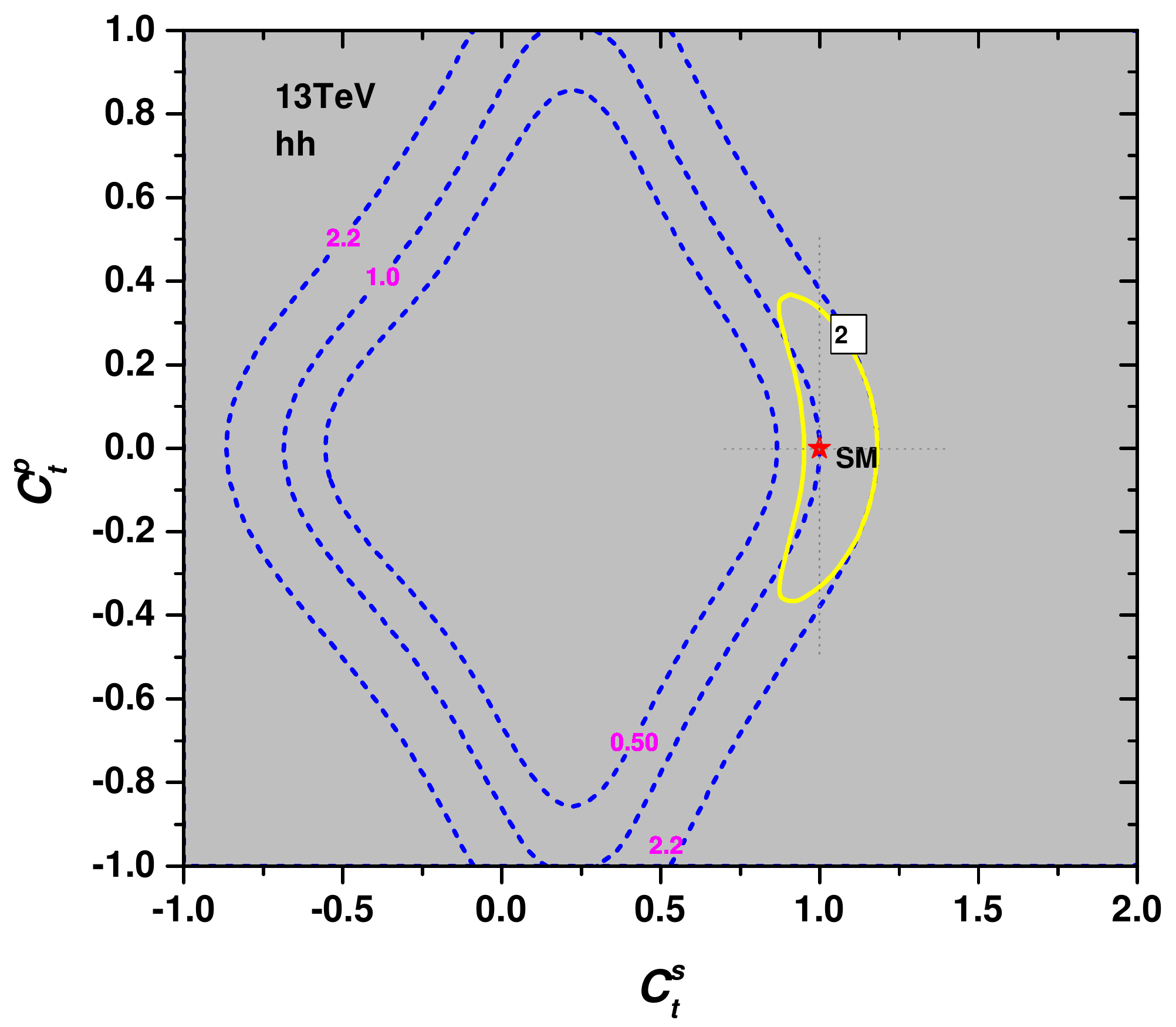}\vspace{-0.5cm}
\caption{ The signal strength $\mu_{t\bar{t}h,~thj,~hh}$ at 13 TeV LHC on the plane of $C^{s}_t$ and $C^{p}_t$. The yellow contour corresponds to the allowed $2\sigma$ limit in Fig~\ref{fit}.}
\label{lhc}
\end{figure}
In Fig.~\ref{lhc}, we show the signal strength $\mu_{t\bar{t}h,~thj,~hh}$ at 13 TeV LHC on the plane of $C^{s}_t$ and $C^{p}_t$. The yellow contour corresponds to the $2\sigma$ limit from Higgs data fit in Fig~\ref{fit}. From Fig.~\ref{lhc}, we can see that the cross sections of $t\bar{t}h$, $thj$ and $hh$ can be enhanced up to 1.41, 1.18 and 2.2 times as large as the SM predictions respectively within the current allowed parameter space. For $C^p_t=0$ and $C^s_t>1$, the cross section of the processes $pp \to t\bar{t}h,~tjh,~hh$ becomes larger with the increase of $C^s_t$. For $C^s_t=1$, the presence of the pseudo-component $C^p_t$ will enhance the cross sections of all these Higgs production processes at the LHC. Besides, the signal strength $\mu_{t\bar{t}h}$ is symmetric around the parameter point $C^s_t=C^p_t=0$, while $\mu_{tjh}$ and $\mu_{hh}$ is asymmetric because the processes $pp \to tjh,~hh$ can be induced without the top-Higgs coupling.

\subsection{$hZ$ and $h\gamma $ production at $e^+e^-$ colliders}
\begin{figure}[ht]
\centering
\includegraphics[width=3.2in,height=3in]{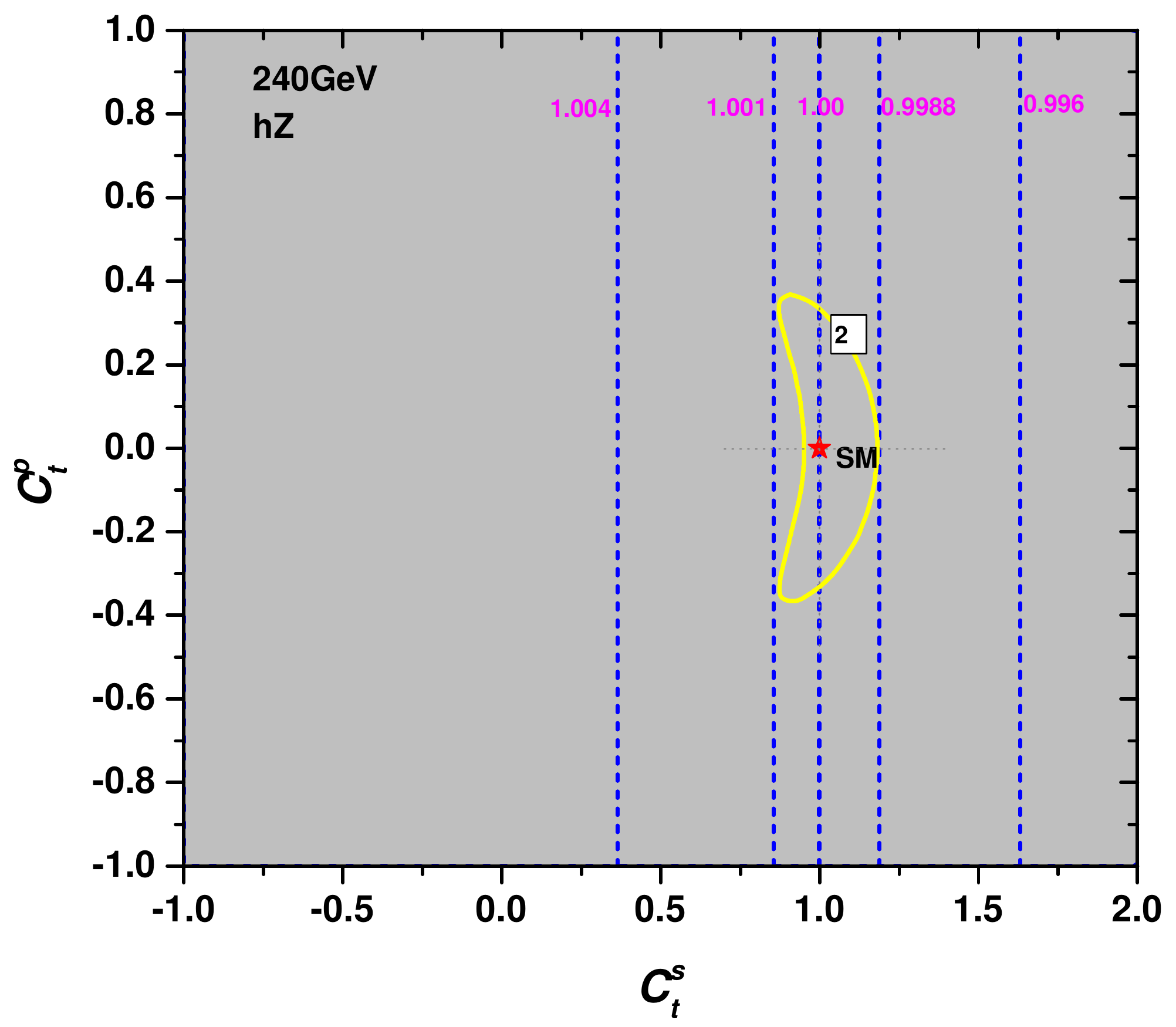}
\includegraphics[width=3.2in,height=3in]{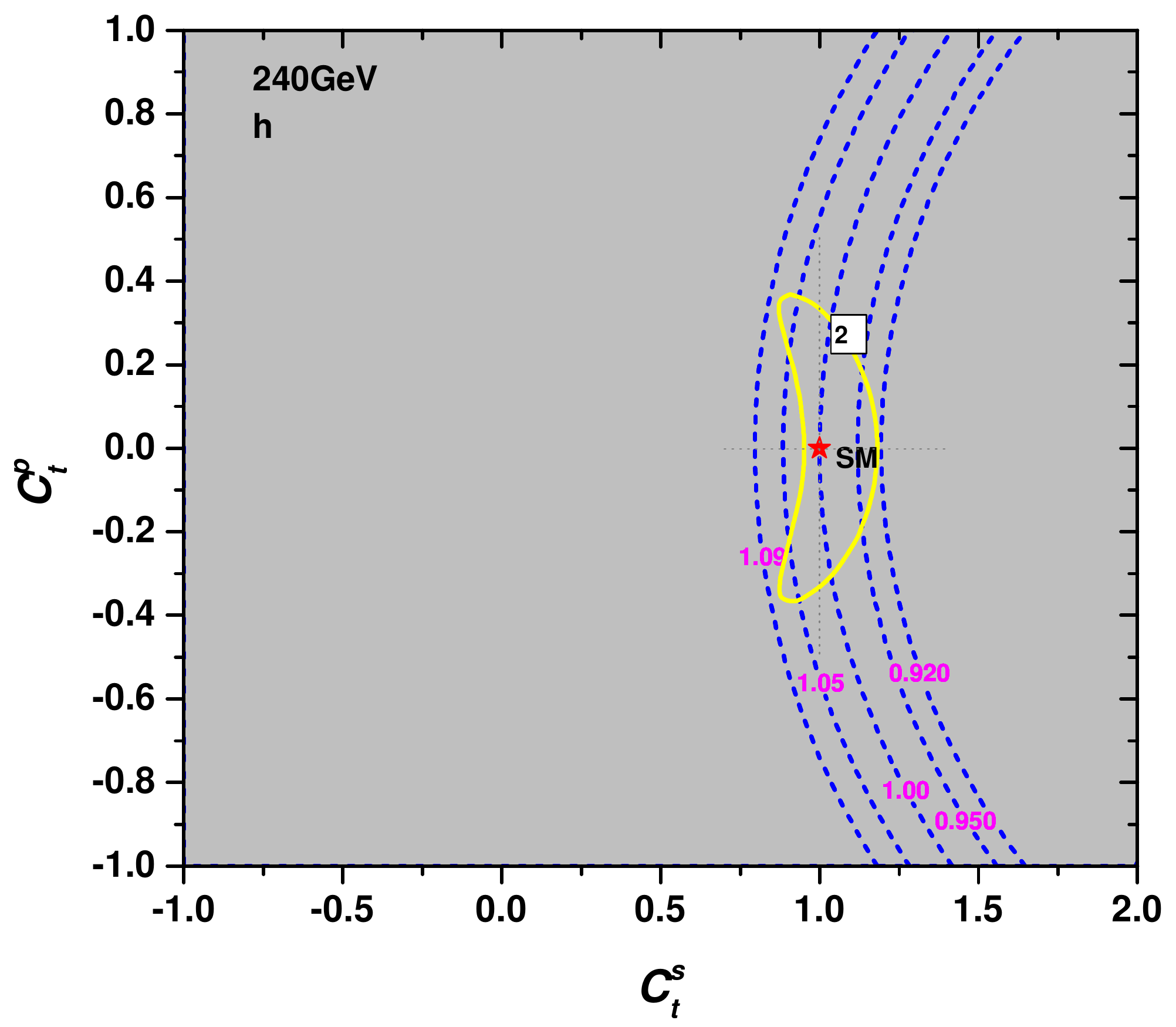}\vspace{-0.5cm}
\caption{Same as Fig.~\ref{lhc}, but for the signal strength $\mu_{hZ,~h\gamma}$ at future $e^+e^-$ colliders with $\sqrt{s}=240$ GeV.}
\label{cepc}
\end{figure}
In Fig.~\ref{cepc}, we present the signal strength $\mu_{hZ,~h\gamma}$ at a $e^+e^-$ collider with $\sqrt{s}=240$ GeV on the plane of $C^{s}_t$ and $C^{p}_t$. It is seen here that the CP violating top-Higgs couplings can enhance the SM cross sections of $e^+e^- \to hZ$, and $e^+e^- \to h\gamma$ by 0.1\% and 9\% respectively within the current allowed parameter space. For $C^p_t=0$ and $C^s_t>1$, the cross section of the process $e^+e^-\to hZ$ becomes larger with the increase of $C^s_t$, while the cross section of the process $e^+e^- \to h\gamma$ decreases because of the cancelation between top quark loop and $W$ boson loop. It should be noted that the presence of the pseudo-component $C^p_t$ only enhances the cross section of $e^+e^- \to h\gamma$. The cross section of $e^+e^- \to hZ$ is independent of $C^p_t$. The future precision measurement of the process $e^+e^- \to h\gamma$ with an accuracy of 5\% will be able to constrain $|C^p_t|<0.19$ at most at a 240 GeV $e^+e^-$ Higgs factory.

\section{conclusions}\label{section4}
In this paper, we updated the bound on the CP-violating top-Higgs couplings by combining the LHC Run-1 and -2 Higgs data sets. We find that the CP-odd component $C^p_t$ and the CP-even component $C^s_t$ have been constrained within the ranges $|C^p_t|< 0.37$ and $0.85 <C^s_t< 1.20$ at $2\sigma$ level, which is stronger than the previous LHC Run-1 bound $|C^p_t|< 0.54$ and $0.68 <C^s_t< 1.20$. Under the new bound, we examine the impact of the CP-violating top-Higgs couplings in the Higgs production processes $pp \to t\bar{t}h, thj, hh$ at 13 TeV LHC, and $e^+e^- \to Zh, h\gamma$ at a future 240 GeV Higgs factory. We note that the cross sections of $pp \to t\bar{t}h$, $pp \to thj$, $pp \to hh$, $e^+e^- \to hZ$ and $e^+e^- \to h\gamma$ can be still enhanced up to 1.41, 1.18, 2.2, 1.001 and 1.09 times as large as the SM predictions, respectively. The future precision measurement of the process $e^+e^- \to h\gamma$ with an accuracy of 5\% will be able to constrain $|C^p_t|<0.19$ at most at a 240 GeV $e^+e^-$ Higgs factory.

\acknowledgments
This work is partly supported by the Australian Research Council, by China Scholarship Council, by the National Natural Science Foundation of China (NNSFC) under grants Nos. 11275057, 11305049 and 11405047, by Specialised Research Fund for the Doctoral Program of Higher Education under Grant No. 20134104120002 and by the Startup Foundation for Doctors of Henan Normal University under contract No.11112.


\begin{thebibliography}{99}
\bibitem{higgs-atlas}
G. Aad et al.(ATLAS Collaboration), Phys. Lett. B \textbf{710}, 49
(2012).

\bibitem{higgs-cms}
S. Chatrachyan et al.(CMS Collaboration), Phys. Lett. B
\textbf{710}, 26 (2012).



\bibitem{Coleman:1969sm}
  S.~R.~Coleman, J.~Wess and B.~Zumino,
  Phys.\ Rev.\  {\bf 177}, 2239 (1969).
  doi:10.1103/PhysRev.177.2239

\bibitem{Callan:1969sn}
  C.~G.~Callan, Jr., S.~R.~Coleman, J.~Wess and B.~Zumino,
  Phys.\ Rev.\  {\bf 177}, 2247 (1969).
  doi:10.1103/PhysRev.177.2247

\bibitem{Sher:1988mj}
  M.~Sher,
  Phys.\ Rept.\  {\bf 179}, 273 (1989).
  doi:10.1016/0370-1573(89)90061-6

\bibitem{Degrassi:2012ry}
  G.~Degrassi, S.~Di Vita, J.~Elias-Miro, J.~R.~Espinosa, G.~F.~Giudice, G.~Isidori and A.~Strumia,
  JHEP {\bf 1208}, 098 (2012)
  doi:10.1007/JHEP08(2012)098
  [arXiv:1205.6497 [hep-ph]].


\bibitem{Zhang:1994fb}
  X.~Zhang, S.~K.~Lee, K.~Whisnant and B.~L.~Young,
  Phys.\ Rev.\ D {\bf 50}, 7042 (1994)
  doi:10.1103/PhysRevD.50.7042
  [hep-ph/9407259].

\bibitem{Kobakhidze:2015xlz}
  A.~Kobakhidze, L.~Wu and J.~Yue,
  JHEP {\bf 1604}, 011 (2016)
  doi:10.1007/JHEP04(2016)011
  [arXiv:1512.08922 [hep-ph]].

\bibitem{Huang:2015izx}
  F.~P.~Huang, P.~H.~Gu, P.~F.~Yin, Z.~H.~Yu and X.~Zhang,
  Phys.\ Rev.\ D {\bf 93}, no. 10, 103515 (2016)
  doi:10.1103/PhysRevD.93.103515
  [arXiv:1511.03969 [hep-ph]].


\bibitem{Kobakhidze:2012wb}
  A.~Kobakhidze,
  arXiv:1208.5180 [hep-ph].


\bibitem{Cheung:2014noa}
  K.~Cheung, J.~S.~Lee and P.~Y.~Tseng,
  Phys.\ Rev.\ D {\bf 90}, 095009 (2014)
  doi:10.1103/PhysRevD.90.095009
  [arXiv:1407.8236 [hep-ph]].

\bibitem{Kobakhidze:2014gqa}
  A.~Kobakhidze, L.~Wu and J.~Yue,
  JHEP {\bf 1410}, 100 (2014)
  doi:10.1007/JHEP10(2014)100
  [arXiv:1406.1961 [hep-ph]].

\bibitem{Dolan:2014upa}
  M.~J.~Dolan, P.~Harris, M.~Jankowiak and M.~Spannowsky,
  Phys.\ Rev.\ D {\bf 90}, 073008 (2014)
  doi:10.1103/PhysRevD.90.073008
  [arXiv:1406.3322 [hep-ph]].

\bibitem{Khatibi:2014bsa}
  S.~Khatibi and M.~Mohammadi Najafabadi,
  Phys.\ Rev.\ D {\bf 90}, no. 7, 074014 (2014)
  doi:10.1103/PhysRevD.90.074014
  [arXiv:1409.6553 [hep-ph]].

\bibitem{Chien:2015xha}
  Y.~T.~Chien, V.~Cirigliano, W.~Dekens, J.~de Vries and E.~Mereghetti,
  JHEP {\bf 1602}, 011 (2016)
  [JHEP {\bf 1602}, 011 (2016)]
  doi:10.1007/JHEP02(2016)011
  [arXiv:1510.00725 [hep-ph]].

\bibitem{Cirigliano:2016njn}
  V.~Cirigliano, W.~Dekens, J.~de Vries and E.~Mereghetti,
  Phys.\ Rev.\ D {\bf 94}, no. 1, 016002 (2016)
  doi:10.1103/PhysRevD.94.016002
  [arXiv:1603.03049 [hep-ph]].


\bibitem{Cirigliano:2016nyn}
  V.~Cirigliano, W.~Dekens, J.~de Vries and E.~Mereghetti,
  Phys.\ Rev.\ D {\bf 94}, no. 3, 034031 (2016)
  doi:10.1103/PhysRevD.94.034031
  [arXiv:1605.04311 [hep-ph]].

\bibitem{Dolan:2012ac}
  M.~J.~Dolan, C.~Englert and M.~Spannowsky,
  Phys.\ Rev.\ D {\bf 87}, no. 5, 055002 (2013)
  doi:10.1103/PhysRevD.87.055002
  [arXiv:1210.8166 [hep-ph]].


\bibitem{Nishiwaki:2013cma}
  K.~Nishiwaki, S.~Niyogi and A.~Shivaji,
  JHEP {\bf 1404}, 011 (2014)
  doi:10.1007/JHEP04(2014)011
  [arXiv:1309.6907 [hep-ph]].

\bibitem{Han:2013sga}
  C.~Han, X.~Ji, L.~Wu, P.~Wu and J.~M.~Yang,
  JHEP {\bf 1404}, 003 (2014)
  doi:10.1007/JHEP04(2014)003
  [arXiv:1307.3790 [hep-ph]].

\bibitem{Liu:2014rba}
  N.~Liu, S.~Hu, B.~Yang and J.~Han,
  JHEP {\bf 1501}, 008 (2015)
  doi:10.1007/JHEP01(2015)008
  [arXiv:1408.4191 [hep-ph]].

\bibitem{Goertz:2014qta}
  F.~Goertz, A.~Papaefstathiou, L.~L.~Yang and J.~Zurita,
  JHEP {\bf 1504}, 167 (2015)
  doi:10.1007/JHEP04(2015)167
  [arXiv:1410.3471 [hep-ph]].


\bibitem{Dawson:2015oha}
  S.~Dawson, A.~Ismail and I.~Low,
  Phys.\ Rev.\ D {\bf 91}, no. 11, 115008 (2015)
  doi:10.1103/PhysRevD.91.115008
  [arXiv:1504.05596 [hep-ph]].

\bibitem{Shen:2015pha}
  C.~Shen and S.~h.~Zhu,
  Phys.\ Rev.\ D {\bf 92}, no. 9, 094001 (2015)
  doi:10.1103/PhysRevD.92.094001
  [arXiv:1504.05626 [hep-ph]].

\bibitem{Wu:2015nba}
  L.~Wu, J.~M.~Yang, C.~P.~Yuan and M.~Zhang,
  Phys.\ Lett.\ B {\bf 747}, 378 (2015)
  doi:10.1016/j.physletb.2015.06.020
  [arXiv:1504.06932 [hep-ph]].

\bibitem{Lu:2015jza}
  C.~T.~Lu, J.~Chang, K.~Cheung and J.~S.~Lee,
  JHEP {\bf 1508}, 133 (2015)
  doi:10.1007/JHEP08(2015)133
  [arXiv:1505.00957 [hep-ph]].

\bibitem{Cao:2015oaa}
  Q.~H.~Cao, B.~Yan, D.~M.~Zhang and H.~Zhang,
  Phys.\ Lett.\ B {\bf 752}, 285 (2016)
  doi:10.1016/j.physletb.2015.11.045
  [arXiv:1508.06512 [hep-ph]].

\bibitem{Marciano:1991qq}
  W.~J.~Marciano and F.~E.~Paige,
  Phys.\ Rev.\ Lett.\  {\bf 66}, 2433 (1991).
  doi:10.1103/PhysRevLett.66.2433

\bibitem{Frederix:2011zi}
  R.~Frederix, S.~Frixione, V.~Hirschi, F.~Maltoni, R.~Pittau and P.~Torrielli,
  Phys.\ Lett.\ B {\bf 701}, 427 (2011)
  doi:10.1016/j.physletb.2011.06.012
  [arXiv:1104.5613 [hep-ph]].


\bibitem{Degrande:2012gr}
  C.~Degrande, J.~M.~Gerard, C.~Grojean, F.~Maltoni and G.~Servant,
  JHEP {\bf 1207}, 036 (2012)
  Erratum: [JHEP {\bf 1303}, 032 (2013)]
  doi:10.1007/JHEP07(2012)036, 10.1007/JHEP03(2013)032
  [arXiv:1205.1065 [hep-ph]].

\bibitem{Liu:2015aka}
  N.~Liu, Y.~Zhang, J.~Han and B.~Yang,
  JHEP {\bf 1509}, 008 (2015)
  doi:10.1007/JHEP09(2015)008
  [arXiv:1503.08537 [hep-ph]].

\bibitem{Buckley:2015vsa}
  M.~R.~Buckley and D.~Goncalves,
  Phys.\ Rev.\ Lett.\  {\bf 116}, no. 9, 091801 (2016)
  doi:10.1103/PhysRevLett.116.091801
  [arXiv:1507.07926 [hep-ph]].

\bibitem{Cao:2016wib}
  Q.~H.~Cao, S.~L.~Chen and Y.~Liu,
  arXiv:1602.01934 [hep-ph].

\bibitem{Maltoni:2016yxb}
  F.~Maltoni, E.~Vryonidou and C.~Zhang,
  arXiv:1607.05330 [hep-ph].

\bibitem{Gritsan:2016hjl}
  A.~V.~Gritsan, R.~Rontsch, M.~Schulze and M.~Xiao,
  Phys.\ Rev.\ D {\bf 94}, no. 5, 055023 (2016)
  doi:10.1103/PhysRevD.94.055023
  [arXiv:1606.03107 [hep-ph]].

\bibitem{Dolan:2016qvg}
  M.~J.~Dolan, M.~Spannowsky, Q.~Wang and Z.~H.~Yu,
  Phys.\ Rev.\ D {\bf 94}, no. 1, 015025 (2016)
  doi:10.1103/PhysRevD.94.015025
  [arXiv:1606.00019 [hep-ph]].

\bibitem{Chang:2016mso}
  J.~Chang, K.~Cheung, J.~S.~Lee and C.~T.~Lu,
  arXiv:1607.06566 [hep-ph].

\bibitem{Maltoni:2001hu}
  F.~Maltoni, K.~Paul, T.~Stelzer and S.~Willenbrock,
  Phys.\ Rev.\ D {\bf 64}, 094023 (2001)
  doi:10.1103/PhysRevD.64.094023
  [hep-ph/0106293].

\bibitem{Lu:2010zzb}
  G.~R.~Lu and L.~Wu,
  Chin.\ Phys.\ Lett.\  {\bf 27}, 031401 (2010).
  doi:10.1088/0256-307X/27/3/031401

\bibitem{Farina:2012xp}
  M.~Farina, C.~Grojean, F.~Maltoni, E.~Salvioni and A.~Thamm,
  JHEP {\bf 1305}, 022 (2013)
  doi:10.1007/JHEP05(2013)022
  [arXiv:1211.3736 [hep-ph]].

\bibitem{Biswas:2012bd}
  S.~Biswas, E.~Gabrielli and B.~Mele,
  JHEP {\bf 1301}, 088 (2013)
  doi:10.1007/JHEP01(2013)088
  [arXiv:1211.0499 [hep-ph]].


\bibitem{Ellis:2013yxa}
  J.~Ellis, D.~S.~Hwang, K.~Sakurai and M.~Takeuchi,
  JHEP {\bf 1404}, 004 (2014)
  doi:10.1007/JHEP04(2014)004
  [arXiv:1312.5736 [hep-ph]].

\bibitem{Englert:2014pja}
  C.~Englert and E.~Re,
  Phys.\ Rev.\ D {\bf 89}, no. 7, 073020 (2014)
  doi:10.1103/PhysRevD.89.073020
  [arXiv:1402.0445 [hep-ph]].

\bibitem{Chang:2014rfa}
  J.~Chang, K.~Cheung, J.~S.~Lee and C.~T.~Lu,
  JHEP {\bf 1405}, 062 (2014)
  doi:10.1007/JHEP05(2014)062
  [arXiv:1403.2053 [hep-ph]].


\bibitem{Wu:2014dba}
  L.~Wu,
  JHEP {\bf 1502}, 061 (2015)
  doi:10.1007/JHEP02(2015)061
  [arXiv:1407.6113 [hep-ph]].

\bibitem{Yang:2014xma}
  B.~Yang, J.~Han and N.~Liu,
  JHEP {\bf 1504}, 148 (2015)
  doi:10.1007/JHEP04(2015)148
  [arXiv:1412.2927 [hep-ph]].

\bibitem{Yue:2014tya}
  J.~Yue,
  Phys.\ Lett.\ B {\bf 744}, 131 (2015)
  doi:10.1016/j.physletb.2015.03.044
  [arXiv:1410.2701 [hep-ph]].

\bibitem{Rindani:2016scj}
  S.~D.~Rindani, P.~Sharma and A.~Shivaji,
  Phys.\ Lett.\ B {\bf 761}, 25 (2016)
  doi:10.1016/j.physletb.2016.08.002
  [arXiv:1605.03806 [hep-ph]].

\bibitem{Liu:2016dag}
  Y.~B.~Liu and Z.~J.~Xiao,
  Phys.\ Rev.\ D {\bf 94}, no. 5, 054018 (2016)
  doi:10.1103/PhysRevD.94.054018
  [arXiv:1605.01179 [hep-ph]].


\bibitem{Hagiwara:1989xx}
  K.~Hagiwara and H.~Murayama,
  Phys.\ Rev.\ D {\bf 41}, 1001 (1990).
  doi:10.1103/PhysRevD.41.1001

\bibitem{Kniehl:1995tn}
  B.~A.~Kniehl and M.~Spira,
  Z.\ Phys.\ C {\bf 69}, 77 (1995)
  doi:10.1007/s002880050007
  [hep-ph/9505225].


\bibitem{Hu:2014eia}
  S.~L.~Hu, N.~Liu, J.~Ren and L.~Wu,
  J.\ Phys.\ G {\bf 41}, no. 12, 125004 (2014)
  doi:10.1088/0954-3899/41/12/125004
  [arXiv:1402.3050 [hep-ph]].

\bibitem{Ren:2015uka}
  H.~Y.~Ren,
  Chin.\ Phys.\ C {\bf 39}, no. 11, 113101 (2015)
  doi:10.1088/1674-1137/39/11/113101
  [arXiv:1503.08307 [hep-ph]].

\bibitem{Cao:2015iua}
  Q.~H.~Cao, H.~R.~Wang and Y.~Zhang,
  Chin.\ Phys.\ C {\bf 39}, no. 11, 113102 (2015)
  doi:10.1088/1674-1137/39/11/113102
  [arXiv:1505.00654 [hep-ph]].

\bibitem{Li:2015kxc}
  G.~Li, H.~R.~Wang and S.~h.~Zhu,
  Phys.\ Rev.\ D {\bf 93}, no. 5, 055038 (2016)
  doi:10.1103/PhysRevD.93.055038
  [arXiv:1506.06453 [hep-ph]].

\bibitem{AguilarSaavedra:2009mx}
  J.~A.~Aguilar-Saavedra,
  Nucl.\ Phys.\ B {\bf 821}, 215 (2009)
  doi:10.1016/j.nuclphysb.2009.06.022
  [arXiv:0904.2387 [hep-ph]].

\bibitem{Brod:2013cka}
  J.~Brod, U.~Haisch and J.~Zupan,
  JHEP {\bf 1311}, 180 (2013)
  doi:10.1007/JHEP11(2013)180
  [arXiv:1310.1385 [hep-ph], arXiv:1310.1385].


\bibitem{Barr:1990vd}
  S.~M.~Barr and A.~Zee,
  Phys.\ Rev.\ Lett.\  {\bf 65}, 21 (1990)
  Erratum: [Phys.\ Rev.\ Lett.\  {\bf 65}, 2920 (1990)].
  doi:10.1103/PhysRevLett.65.21

\bibitem{Stockinger:2006zn}
  D.~Stockinger,
  J.\ Phys.\ G {\bf 34}, R45 (2007)
  doi:10.1088/0954-3899/34/2/R01
  [hep-ph/0609168].

\bibitem{Baron:2013eja}
  J.~Baron {\it et al.} [ACME Collaboration],
  Science {\bf 343}, 269 (2014)
  doi:10.1126/science.1248213
  [arXiv:1310.7534 [physics.atom-ph]].



\bibitem{Bernon:2015hsa}
  J.~Bernon and B.~Dumont,
  Eur.\ Phys.\ J.\ C {\bf 75}, no. 9, 440 (2015)
  doi:10.1140/epjc/s10052-015-3645-9
  [arXiv:1502.04138 [hep-ph]].

\bibitem{ATLAS-CONF-2016-068}
ATLAS-CONF-2016-068.

\bibitem{ATLAS-CONF-2016-081}
ATLAS-CONF-2016-081.

\bibitem{ATLAS-CONF-2016-091}
ATLAS-CONF-2016-068.

\bibitem{CMS-PAS-HIG-16-020}
CMS-PAS-HIG-16-020.

\bibitem{CMS-PAS-HIG-16-033}
CMS-PAS-HIG-16-033.

\bibitem{CMS-PAS-HIG-16-022}
CMS-PAS-HIG-16-022.

\bibitem{pdg}
J. Beringer {\it et al.}, Particle Data Group, Phys.\ Rev. \ D {\bf
86}, 010001 (2012) and 2013 partial update for the 2014 edition.

\bibitem{cteq}
P.~M.~Nadolsky, H.~-L.~Lai, Q.~-H.~Cao, J.~Huston, J.~Pumplin,
D.~Stump, W.~-K.~Tung and C.~-P.~Yuan,
Phys.\ Rev.\ D {\bf 78}, 013004 (2008), arXiv:0802.0007 [hep-ph].

\bibitem{feynarts}
T. Hahn, Comput. Phys. Commun. \textbf{140}, 418 (2001).


\bibitem{formcalc}
T. Hahn, M. Perez-Victoria, Comput. Phys. Commun. \textbf{118}, 153
(1999).

\bibitem{looptools}
G. J. van Oldenborgh, Phys Commun \textbf{66}, 1 (1991).

\bibitem{two-cutoff}
  B.~W.~Harris and J.~F.~Owens,
  Phys.\ Rev.\ D {\bf 65}, 094032 (2002)
  doi:10.1103/PhysRevD.65.094032
  [hep-ph/0102128].

\bibitem{ningliu-1}
  N.~Liu,
  Phys.\ Lett.\ B {\bf 707}, 137 (2012)
  doi:10.1016/j.physletb.2011.12.032
  [arXiv:1112.3702 [hep-ph]].

\bibitem{ningliu-2}
  N.~Liu, J.~Ren and B.~Yang,
  Phys.\ Lett.\ B {\bf 731}, 70 (2014)
  doi:10.1016/j.physletb.2014.02.023
  [arXiv:1310.6192 [hep-ph]].

\bibitem{ningliu-3}
  N.~Liu, J.~Ren, L.~Wu, P.~Wu and J.~M.~Yang,
  JHEP {\bf 1404}, 189 (2014)
  doi:10.1007/JHEP04(2014)189
  [arXiv:1311.6971 [hep-ph]].


\end{thebibliography}
\end{document}